\documentclass[iop]{emulateapj}

\usepackage{hyperref}

\slugcomment{The Astrophysical Journal}
\accepted{6 December 2012}

\shorttitle{Proton anisotropy and magnetic reconnection in the solar wind}
\shortauthors{Matteini et al.}

\begin{document}


\title{Proton temperature anisotropy and magnetic reconnection in the solar wind: effects of kinetic instabilities on current sheet stability}


\author{L. Matteini\altaffilmark{1,2}, S. Landi\altaffilmark{1}, M. Velli\altaffilmark{3,1}, and W. H. Matthaeus\altaffilmark{4}}\email{matteini@arcetri.astro.it}
\altaffiltext{1}{Dipartimento di Fisica e Astronomia, Universit\`a degli Studi di Firenze, Largo E. Fermi 2, 50125 Florence, Italy}
\altaffiltext{2}{Imperial College London, London, SW7 2AZ, UK}
\altaffiltext{3}{Jet Propulsion Laboratory, California Institute of Technology, Pasadena, CA, USA}
\altaffiltext{4}{Bartol Research Institute, Department of Physics and Astronomy, University of Delaware, Delaware 19716, USA}


\begin{abstract}
\small{We investigate the role of kinetic instabilities driven by a proton anisotropy on the onset of magnetic reconnection by means of 2-D hybrid simulations. The collisionless tearing of a current sheet is studied in the presence of a proton temperature anisotropy in the surrounding plasma. Our results confirm that anisotropic protons within the current sheet region can significantly enhance/stabilize the tearing instability of the current. Moreover, fluctuations associated to linear instabilities excited by large proton temperature anisotropies can significantly influence the stability of the plasma and perturb the current sheets, triggering the tearing instability. 
We find that such a complex coupling leads to a faster tearing evolution in the $T_\perp>T_\|$ regime when an ion-cyclotron instability is generated by the anisotropic proton distribution functions. On the contrary, in the presence of the opposite anisotropy, fire hose fluctuations excited by the unstable background protons with $T_\|<T_\perp$ are not able to efficiently destabilize current sheets, which remain stable for a long time after fire hose saturation. We discuss possible influences of this novel coupling on the solar wind and heliospheric plasma dynamics.}

\end{abstract}


\keywords{\small Methods: numerical $-$ Magnetic reconnection $-$ Instabilities $-$ Plasmas $-$ Solar wind}


\section{Introduction}
 Particle distribution functions that are not at thermal equilibrium and depart from Maxwellians are frequently observed in space and  astrophysical collisionless plasmas. 
In the weakly collisional solar wind, temperature anisotropies ($T_\perp\ne T_\|$), defined with respect to the local mean magnetic field, are ubiquitous \citep{Marsch_al_1982a}.
In this framework, 
microphysics processes generated by nonthermal particle distributions can importantly affect and change the macroscopic evolution of the system \citep[see for example][for a solar wind review]{Matteini_al_2011}.
Direct measurements of the ion temperatures in the solar wind \citep{Hellinger_al_2006,Matteini_al_2007} and in the magnetosphere \citep{Samsonov_al_2007} suggest that kinetic instabilities driven by a proton temperature anisotropy play a role in constraining the ratio of parallel and perpendicular temperatures.
 Such processes act generating unstable fluctuations that interact with particles and scatter them towards a more isotropic state. Signatures of enhanced wave power associated to kinetic instabilities have been observed in the solar wind \citep{Bale_al_2009, Wicks_al_2012}, supporting the idea that these mechanisms are active in the plasma expansion.
Observations suggest that similar instabilities are also at work for the other solar wind plasma components, as alpha particles \citep{Maruca_al_2012}, and possibly electrons \citep{Stverak_al_2008}, despite for the latter also Coulomb collisions are expected to significantly play a role \citep{Landi_al_2012}.
 These kinetic processes are believed to be relevant also in astrophysical plasmas, as galaxy clusters \citep{Schekochihin_al_2005, Schekochihin_al_2010, Kunz_al_2011} and accretion discs \citep{Sharma_al_2007, Riquelme_al_2012}. Then their investigation in the solar wind, where thanks to the in situ measurements by spacecrafts we have access to direct comparisons between theory and observations, is important also for the interpretation and modeling of other astrophysical systems.

At the same time, discontinuities of the magnetic field are often observed in the solar wind
\citep[e.g.,][]{Erdos_Balogh_2008}. These current sheet structures are possible sites of magnetic reconnection \citep[e.g.,][]{Servidio_al_2011}, and seem to be associated with local plasma heating \citep{Osman_al_2012b}.
The role of reconnection in the solar wind is however not fully understood yet, and identified reconnection events constitute only a fraction of the total observed discontinuities in the solar wind \citep[e.g.,][]{Gosling_2007}. At the same time, its contribution to the solar wind heating problem is still an open issue; some authors \citep{Borovsky_Denton_2011} have found a lack of signatures of preferential plasma heating correlated with strong current sheets. These observations suggest that other properties than exclusively the thickness of the current sheets may play a role in determining the stability of magnetic structures and controlling their dissipation through reconnection processes.
A recent observational analysis \citep{Osman_al_2012a} has shown that the statistical distribution of events associated to current sheets in the solar wind display some possible correlations with the values of the proton temperature anisotropy, motivating then new modeling and theoretical studies of the coupling between kinetic processes, as the instabilities described above, and solar wind structures.
Usually wave-particle interactions and structures are considered independently, however in non homogeneous plasmas particle distribution anisotropies may have an important effects on the stability of gradients associated with current sheets, velocity shears and density.
It is then an open question how small scale processes develop in non homogeneous plasmas and how they interact with coherent structures that are present in the system.

It is known that in collisionless systems current sheets are unstable against tearing instability, a process where the current tends to collapse into filaments. 
The tearing instability produces magnetic islands that then interact and merge together giving rise to a nonlinear instability phase, where the reconnection process is enhanced.
It has recently pointed out that reconnection processes can significantly influence the local particle distributions; temperature anisotropies associated to reconnecting regions are observed in numerical simulations \citep{Aunai_al_2011, Servidio_al_2012}.
Kinetic instabilities generated by these anisotropic distributions have been suggested to be efficient mechanisms able to constrain the size of magnetic islands during reconnection events \citep{Schoeffler_al_2011} and to play a role in the associated particle acceleration \citep{Drake_al_2010}.

On the other hand, the role of temperature anisotropy and of associated driven instabilities on the onset of magnetic reconnection has been less investigated.
Some authors \citep[e.g.][]{Chen_Davidson_1981, Coppi_1983, Quest_al_2010} have addressed this problem in the framework of linear theory.
In particular \cite{Chen_Palmadesso_1984} have predicted that in the presence of a $T_\perp>T_\|$ proton anisotropy, the growth rate of the tearing instability is expected to increase with respect the isotropic case, and that the most unstable mode is shifted to larger wavenumbers. On the other side, they also suggested that the tearing instability can be strongly suppressed if the opposite anisotropy $T_\perp<T_\|$ is present in the current sheet.
This picture has been confirmed by numerical studies in a hybrid \citep{Ambrosiano_al_1986} and fluid \citep{Shi_al_1987} framework. 

The aim of this paper is to discuss the stability of small scale magnetic structures related to current sheets in plasmas that are not at thermal equilibrium, as in the case of the solar wind and heliospheric plasma.  
Our analysis constitutes a first attempt of coupling in the same numerical model two phenomena, kinetic microinstabilities and magnetic reconnection, which are in general investigated independently.
We report results from 2-D numerical hybrid simulations that describe the development of a tearing instability of the current in the presence of anisotropic protons, with variable temperature ratio.
We find that kinetic instabilities, analogous to those for homogeneous plasma, can develop in the regions away from current sheets and  generate waves that subsequently perturb and enhance the observed reconnection process at the current sheets.  
Following presentation of the results we will discuss possible influences of this novel coupling and its nonlinear evolution on the interplanetary and coronal plasma dynamics.

\section{Simulation results}
\subsection{Numerical setup}
\begin{figure}
\includegraphics[width=8.5cm]{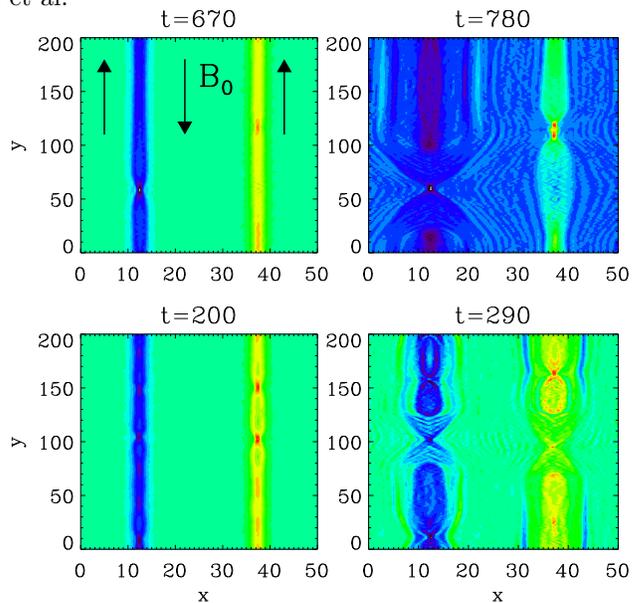}
\caption{Evolution of tearing mode in cases with isotropic proton temperature (top panels) and anisotropic protons in the current sheet region (bottom panels): out of the plane current $J_z$ during the linear (left) and nonlinear (right) phases of the instability. The background plasma is isotropic. Arrows in left top panel indicate the local direction of the ambient magnetic field $B_y(x)=B_0$ in the homogeneous regions around the current sheets.}\label{fig1}  
\end{figure}

We use a two-dimensional hybrid PIC code \citep{Matthews_1994}, describing ions as particles and electrons as a massless charge neutralizing fluid. We initialize the system with a Harris equilibrium model \citep{Harris_1962} for the magnetic field: 
\begin{equation}
B_y(x)=B_0 {\rm tanh}(x/l)
\end{equation}
\begin{equation}
n(x)=n_{cs} {\rm sech^2}(x/l)+n_b
\end{equation}
with a maximum density $n_0=1$ at the current sheet. We add a background proton population with $n_b=0.2n_0$ uniformly in the box.
Units of space and time in the simulations are the ion inertial length $c/\omega_{pi}$ ($\omega_{pi}$ is the plasma frequency) and the inverse of the proton cyclotron frequency, $\Omega_{cp}^{-1}$, respectively.
The adopted simulation box is long  50 $c/\omega_{pi}$ in the $x$ direction and 200 $c/\omega_{pi}$ in the $y$ direction, with a 100x200 grid and spatial resolution $\Delta x=0.5$ $c/\omega_{pi}$ and $\Delta y=1$ $c/\omega_{pi}$; we use $10^3$ particles per cell.  The box has periodic boundaries in both directions and thus contains 2 current sheets of width $l=1.5c/\omega_{pi}$. Note that the spatial separation of the current sheets is such that they do not influence each other during the development of the tearing instability or of the anisotropy driven kinetic instabilities. As expected, interactions between islands associated to distinct sheets are later observed, but only at the final stage of the nonlinear phase of the runs. 
The parallel beta $\beta_\|=8\pi nk_BT_\|/B_0^2$ of the background plasma is set equal to 1 and the electron beta to 0.5.  
The temperature anisotropy ($T_\perp/T_\|$), with respect the equilibrium magnetic field, of both populations is varied in order to test the stability of the magnetic configuration with respect different plasma conditions.
Note that in such a configuration there are regions between the current sheets of homogeneous and constant $B_0$ magnetic field, aligned with $\pm y$ (see top left panel of Figure \ref{fig1}), where fluctuations generated by small scale processes can develop and propagate.

\subsection{Simulations with isotropic background}
We start our analysis with a simulation in which all the plasma is isotropic and we focus on the development of a tearing instability from the initial configuration.
In the top panels of Figure \ref{fig1} we report the evolution of the current sheets at two simulation times. At left, at time $t=670$, at the end of the linear growth a tearing mode with $m=1$ is observed; at a later time, $t=780$, the dynamics becomes nonlinear and a consequent strong deformation of the current is visible in the right panel.
We underline that the instability passes through a long linear phase and highly nonlinear only after $t\sim650$. This assures that, despite the adopted initial conditions are not an exact kinetic equilibrium \citep[e.g.][]{Daughton_1999, Belmont_al_2012}, we do not observe other processes destabilizing the system on a shorter time scale than the classic tearing instability.

\begin{figure}[t]
\includegraphics[width=8.5cm,height=7.5cm]{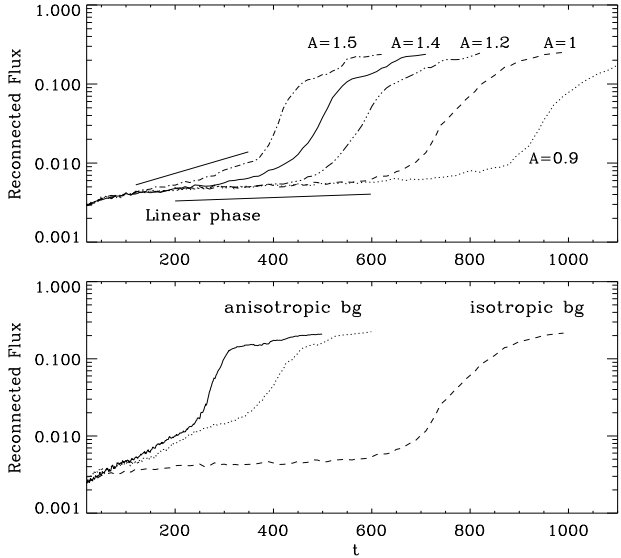}   
\caption{Top panel: Temporal evolution of the reconnected flux for runs with isotropic background and different initial current sheet anisotropy $A=T_\perp/T_\|$. Line encode runs with: $T_\perp=0.9T_\|$ (dotted), $T_\|=T_\perp$ (dashed), $T_\perp=1.2T_\|$ (dash-dot-dot-dotted), $T_\perp=1.4T_\|$ (solid), and $T_\perp=1.5T_\|$ (dash-dotted). Bottom panel: reconnected flux in the presence of unstable anisotropic background protons with $T_\perp=2T_\|$, for a thin ($l=1.5$, solid) and a thicker ($l=3$, dotted) current sheet; the full isotropic plasma case is reported in dashed line as a reference.}\label{fig2}
\end{figure}

We consider now the case of a plasma with a temperature anisotropy.
We have performed simulations using several different initial temperature anisotropies for the current sheet population. At this stage the background proton population is maintained isotropic.
Our investigation qualitatively confirms the picture of \cite{Chen_Palmadesso_1984}, with a faster tearing for larger perpendicular temperature, and a weaker instability if the parallel temperature is larger.
Increasing the anisotropy with larger perpendicular temperatures leads to a shift of the most unstable tearing mode to larger wavenumbers with an increase of the instability growth rate. This is illustrated in the bottom panels of Figure \ref{fig1} that reports the evolution of $J_z$ for a simulation with $T_\perp=2T_\|$ at time $t=200$ (left) and $t=290$ (right). Starting with such an initial anisotropy for the proton population at the current sheet, gives rise to the excitation of a higher mode with $m=6$ and this occurs on a significantly faster timescale compared to the isotropic case (top panels of the figure), leading to the formation of several magnetic islands along the current sheets; subsequently, these are observed to merge giving rise to a strong nonlinear phase (right panel) already at $t\sim290$.

The top panel of Figure \ref{fig2} summarizes the results obtained for various initial temperature anisotropies, reporting the reconnected flux evolution for different runs.  Dotted line refers to the isotropic reference case. From right to left, the lines encode: $T_\perp=0.9T_\|$ (dotted), $T_\|=T_\perp$ (dashed), $T_\perp=1.2T_\|$ (dash-dot-dot-dotted), $T_\perp=1.4T_\|$ (solid), and $T_\perp=1.5T_\|$ (dash-dotted).
Note that in the figure the linear phase of the instability ("tearing") corresponds to the flatter initial part, and that the strong increase of the flux identifies the nonlinear phase, with large amplitude island growth and merging. 
We observe that when increasing the anisotropy (with $T_\perp/T_\|>1$), the linear slope steepens and the nonlinear stage is significantly accelerated. On the other side, when $T_\|>T_\perp$, all the process is delayed.
In agreement with previous studies \citep{Chen_Palmadesso_1984, Ambrosiano_al_1986}, our results confirm that the linear growth rates also depend on the current sheet thickness; we have checked that all the growth rates reported in the figure significantly decrease when taking larger sheet widths.

\begin{figure}
\includegraphics[width=8.5cm]{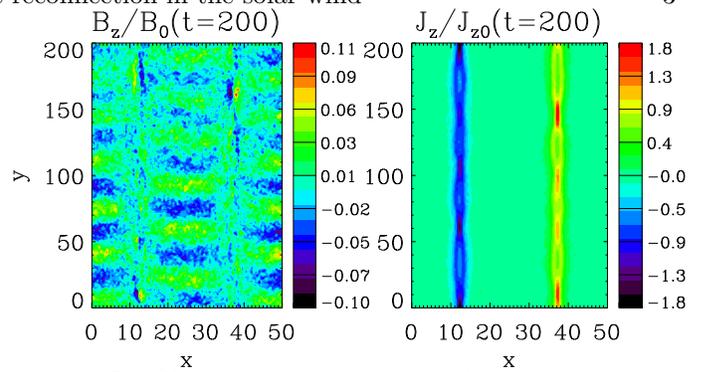}
\caption{
Right panel: $B_z$ component of ion-cyclotron fluctuations generated by unstable background protons with $T_\perp=2T_\|$, at t=200.
Left panel: modulation of the current $J_z$ at the same time.
}
\label{fig3} 
\end{figure}

\subsection{Simulations with anisotropic background}
It is worth noting that when extending the investigation to large temperature ratios the plasma can become unstable with respect kinetic instabilities driven by enhanced thermal anisotropies.
In particular an ion-cyclotron instability can develop  in a $T_\perp>T_\|$ condition \citep[e.g.][]{Gary_al_2003}. This mechanism produces fluctuations that scatter particles and reduce the initial anisotropy bringing the system towards a marginal stable condition.
(Note that a mirror instability can be also excited in the $T_\perp>T_\|$ regime \citep[e.g.][]{Hellinger_2007})
Is then interesting to investigate what happens in the adopted configuration when the injected anisotropy is such to destabilize this linear instability and what are its consequences on the stability of the current sheets. 
To deal with this aspect, we have performed simulations where the temperature anisotropy is not confined in the current sheet region, as in previous section, but it is extended to the whole background plasma.
Note that when this new condition is studied in a regime of stability for the background protons (i.e. weak anisotropy) no peculiar differences are observed in the evolution of the tearing instability with respect the previous cases; the surrounding plasma remains quiet and homogeneous,  because stable, while the current is still destabilized with rates consistent with those reported in top panel of Figure \ref{fig2}, which correspond to an isotropic background plasma.

On the contrary, when the initial anisotropy of the background is such that the uniform regions of plasma are unstable (with $T_\perp=2T_\|$), we observe in the early stages of the simulation the generation of an ion-cyclotron mode, corresponding to the most unstable wavenumber, that starts to propagate along the local magnetic field. Left panel of Figure \ref{fig3} report the out-of-the-plane $B_z$ fluctuations associated to the ion-cyclotron instability which characterize the region of homogeneous magnetic field $B_y$ bounded by the current sheets.
As soon as these are generated, such fluctuations start to perturb the current sheet profiles (right panel); as we will discuss later, thanks to the symmetry imposed by the ion-cyclotron fluctuations at the two sides of the sheet, the forced perturbation consists in a tearing modulation of the current and since this occurs at a scale close to the most unstable current tearing wavenumber, the instability is efficiently triggered. 
As a consequence, this coupling leads to a fast driving of the tearing instability, as shown by the solid line in the bottom panel of Figure \ref{fig2}, that reports the time evolution of the reconnected flux; the dissipation of the current occurs on a shorter time than in the case when the background is isotropic (dashed line).
Moreover, it is remarkable that in the linear part, the growth of the tearing is driven by the rms of the $B_z$ fluctuations excited by the ion-cyclotron instability.
This result demonstrates that unstable fluctuations generated by ion-cyclotron instability importantly influence the stability of the magnetic equilibrium and they can directly accelerate the development of the tearing instability.

It is also important to underline that, while the unforced linear tearing depends strongly on the thickness of the current, on the other side the excitation of fluctuations by kinetic instabilities depends only on the plasma conditions outside the current sheet (temperature anisotropy and beta). As a consequence, the trigger of the tearing instability by ion-cyclotron fluctuations results approximatively independent from the sheet thickness, as shown by the dotted line in the bottom panel of Figure \ref{fig2}, reporting the reconnected flux for a case with a larger  thickness $l=3$, but same proton anisotropy in the background. Despite the increase of sheet thickness and unlike to the classical linear tearing, we still recover a fast tearing, with a growth compatible to that of the thinner sheet case.
Note that the only difference consists in a longer linear phase, since the tearing of the thicker current needs a longer time to become nonlinear.

\begin{figure}
\includegraphics[width=8cm]{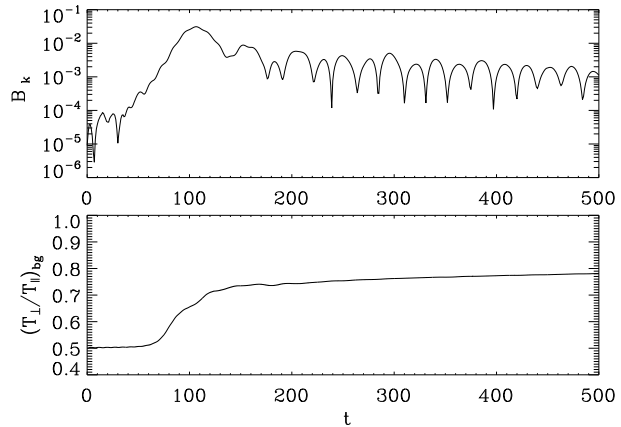}
\caption{Run with initial background anisotropy $T_\|=2T_\perp$ and $\beta_\|=4$, predicted unstable for fire hose instability. Top panel: history of the dominant component of the Fourier transform of $B_z$, associated to a fire hose unstable mode. Bottom panel: evolution of the temperature anisotropy of background protons.}  
\label{fig4}
\end{figure}

An important question at this point concerns whether the presence of any fluctuations induced by a small scale process would be sufficient to trigger such a faster destabilization of the current \citep{Coppi_1983}, and if this may occur independently from the sign of the plasma anisotropy.
To address this issue, we have considered a case with the opposite temperature anisotropy, where, due to a large $T_\|>T_\perp$ anisotropy, protons of the background plasma are predicted unstable with respect a fire hose instability by the linear Vlasov theory. In this situation ($T_\|=2T_\perp$ and $\beta_\|=4$), a strong fire hose mode is soon generated in the box: top panel of Figure \ref{fig4} reports the time history of the $B_z$ Fourier component corresponding to this mode, showing its exponential growth in time for $t<100$. 
The growth rate associated of this mode is in qualitative agreement with the linear Vlasov prediction for homogeneous plasma, even if a deviation is observed, probably due to the more complicated plasma configuration adopted in the simulation.

\begin{figure}[h]
\includegraphics[width=7.9cm]{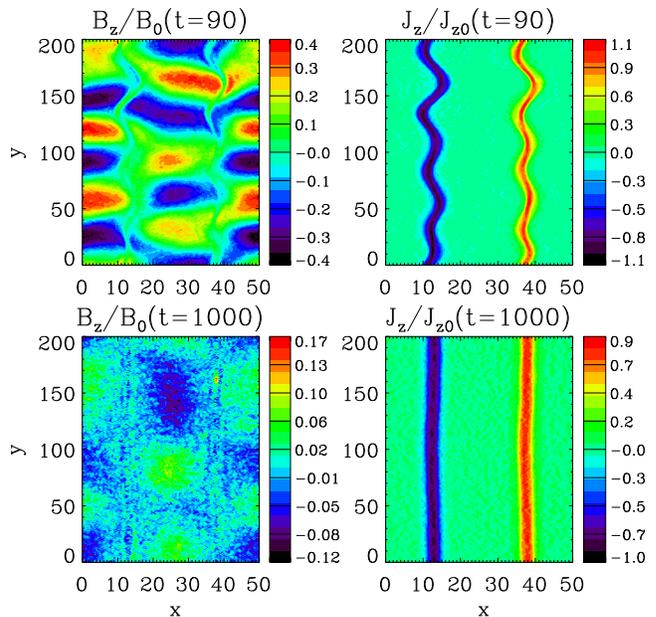}
\caption{Evolution of $B_z$ (left) and $J_z$ (right) for a simulation with anisotropic protons in the background with $T_\|=2T_\perp$ (same run as in Figure \ref{fig4}), during the linear phase of the fire hose instability (top panels) and at a late post-saturation time (bottom).
}
\label{fig5}
\end{figure}

As expected, the effect of this instability is to scatter particles and reduce their temperature anisotropy, driving the plasma to a more isotropic marginal stable state (lower panel). The saturation of the magnetic field fluctuations at $t\gtrsim100$ corresponds to the stabilization of the plasma; after this stage the proton anisotropy remains constant and the fire hose fluctuations are slowly damped in time.
Note that the observed behavior, which is in good agreement with the standard evolution of this instability found in simulations of homogeneous plasma \citep[e.g.][]{Matteini_al_2006}, is here recovered in the case of a non homogeneous plasma, supporting the idea that such processes can be at work in real plasmas, like the solar wind, that are characterized by non thermal distributions and small scale coherent structures.

The activity of the fire hose instability described in the simulation influences and perturbs the magnetic field structure; it is worth noting that the fire hose is an instability of magnetic field oscillations \citep[e.g.][]{Davidson_Volk_1968} and thus leads to an oscillatory modulation of magnetic field lines. As a consequence, its influence on the current sheets is to produce a kink deformation of $J_z$, that becomes more and more significant as the fire hose unstable mode grows and reaches a macroscopic scale before saturation, as shown in the top panels of Figure \ref{fig5}.
It is remarkable that this macroscopic perturbation of the current does not give rise to any tearing instability of the current sheets. On the contrary, these are observed to oscillate passively under the effect of the fire hose activity. Such kink-type oscillations are later damped as soon as, according to Figure \ref{fig4}, the fire hose mode is reabsorbed by the plasma at longer times.
As shown in the bottom panels of Figure \ref{fig5}, at $t=1000$, after the oscillatory stage, the system appears stable again and the current sheets recover their initial equilibrium configuration as at the beginning of the simulation. Starting from this late condition, a tearing instability can then eventually develop and produce the collapse of the current, but only in a significantly longer time than in the cases that are stable against the fire hose. 

\begin{figure}
\includegraphics[width=8.5cm]{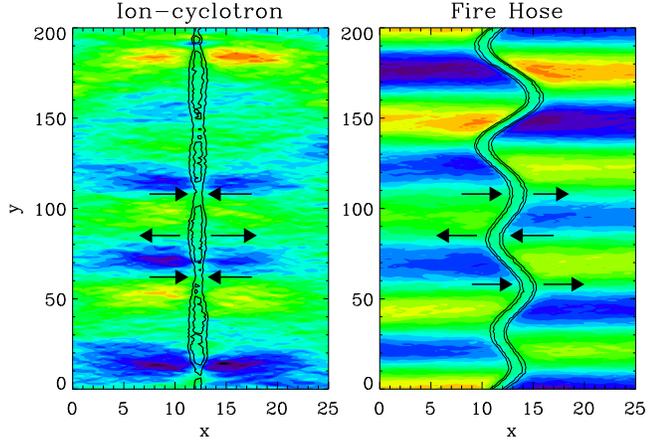}
\caption{The color contour encode the background $B_x$ fluctuations generated by ion-cyclotron (left) and fire hose (right) instabilities. The density of protons at the current sheet at the same simulation time is reported on the superimposed black solid line contour. The arrows show the local direction of the resulting contribution to the electric field component $E_x$ generated by the magnetic fluctuations in the figure.}
\label{fig6}
\end{figure}

To investigate in more detail the origin of the different modulation of the current sheets observed in the simulations, the color contour in Figure \ref{fig5} reports the magnetic fluctuations $B_x$ associated to ion-cyclotron (left) and fire hose (right) instabilities in the background plasma.
The superimposed black solid line contours encode the density of protons within the current sheet. 
Arrows identify the local direction of the electric field generated by the kinetic instability fluctuations.
As shown by the figure, the modulation of the current sheet is very well correlated to these fluctuations, suggesting that the resulting deformation of the current is guided by the background activity, leading to a different evolution according to the different symmetry of the fluctuations. 
The spatial structure of the $x$ component of the magnetic field, $B_x$, is a relevant quantity in our analysis since through the term $J \times B = c/4\pi (\nabla \times B) \times B$
it generates an electric field component pushing particle towards and away from the current sheet. 
In our case this component of $E_x$ is proportional to $(\partial_y B_x)B_0$, which clearly depends on the symmetry of $B_x$ across the discontinuity. When $B_x$ is symmetric through the current sheet (left panel, ion-cyclotron instability case) then the associated electric field has opposite sign on the two sides (due to the orientation change in $B_0$), leading to diverging or converging flows. On the contrary when $B_x$ is antisymmetric (right panel, fire hose case), the electric field does not change orientation across the discontinuity (both $B_x$ and $B_0$ change sign) and gives rise to an oscillation along $y$.
The local direction of this contribution to the electric field is reported by the arrows in the figure and well explains either the kink or the tearing deformation of the current under the effects of the external, anisotropy driven, unstable fluctuations. 

\section{Summary and conclusions}
In summary, we have investigated the properties of current sheets embedded in an anisotropic plasma, focusing on the coupling between small scale processes, temperature anisotropy driven kinetic instabilities, and evolution of plasma structures leading to magnetic reconnection. 
We have tested the stability of an initial Harris equilibrium as a function of the temperature ratio that characterizes protons in the current sheet region and in the background plasma.
In agreement with previous works \citep{Chen_Palmadesso_1984}, we find that the temperature anisotropy plays an important role in determining the stability of the current with respect the tearing instability. 
A $T_\perp>T_\|$ condition within the current sheet leads to faster tearing, and at larger wave numbers, while the opposite $T_\perp<T_\|$ condition has a stabilizing effect.
Moreover, the development of kinetic instabilities that are driven when these temperature anisotropies are distributed in the background plasma, changes further the properties of the system.

It is interesting to note that despite the adopted simulation domain is not homogeneous, as is usually done for kinetic instabilities studies \citep[e.g.][]{Matteini_al_2006}, they are observed to develop anyway; this is because in the box there are regions where the magnetic field is locally uniform (see for example Figure \ref{fig1}) and even if restricted to narrow channels (few inertial lengths), unstable fluctuations can be excited. This suggests that the development of those instabilities, as predicted for a homogeneous Vlasov plasma, is robust and may characterize also the evolution of realistic, non homogeneous systems as space and astrophysical plasmas.

We have shown that ion-cyclotron fluctuations, generated in the presence of an initial unstable $T_\perp>T_\|$ anisotropy in the background
are able to trigger the destabilization of the current and drive a tearing instability  (Figure \ref{fig3})
that emerges more rapidly than it does in the case in which the anisotropy is confined to the current sheet regions and ion-cyclotron modes are not excited. 
Moreover, since such a trigger only depends on the conditions of the background plasma, the tearing instability forced by those fluctuations results to be roughly independent from the current sheet thickness.
On the other hand, we have also shown that fire hose fluctuations that are generated when an excessive $T_\perp<T_\|$ anisotropy characterizes the plasma are not able to trigger enhanced instability of the current. In that case, due to the nature of  those modes (Figure \ref{fig6}), the observed deformation is of the kink-type and does not produce favorable conditions for tearing. As a consequence the current sheets remain stable even when such fluctuations are later dissipated (Figure \ref{fig5}), and remarkably, for a longer time than the isotropic or fire hose stable, case.

We propose that this novel coupling, leading to different evolution according to value of the temperature anisotropy, may take place in the solar wind, where alternatively conditions of $T_\perp>T_\|$ and $T_\perp<T_\|$ for protons are observed \citep{Marsch_al_1982a}. Kinetic instabilities, like fire hose and ion-cyclotron, are believed to play a role in regulating such proton temperature anisotropy along the solar wind expansion \citep{Hellinger_al_2006,Matteini_al_2007,Bale_al_2009} and fluctuations excited by these instabilities can then importantly influence the properties of the plasma. 
Data analysis of small scale fluctuations and particle anisotropies in the solar wind should be complemented with local wind structure informations. This can be relevant in the study of reconnection exhausts as a function of magnetic shear \citep[e.g.][]{Gosling_al_2006}.
We suggest for example that small scale structures, discontinuities and current sheets, may be more observable, because stable, when the plasma is in a $T_\perp<T_\|$ regime, as we have shown that in this regime the plasma is more stable and even external large amplitude fluctuations, like fire hose, are not able to drive the current collapse, in possible agreement with the recent solar wind observations of \cite{Osman_al_2012a}. 
On the ecliptic, this can be also relevant for the dynamics of heliospheric current sheet that is mainly surrounded by slow wind plasma, displaying a majority of $T_\|>T_\perp$ \citep{Hellinger_al_2006}.
On the contrary we expect fast streams observed at high latitude and characterized by protons with highly anisotropic cores with $T_\perp>T_\|$  \citep{Marsch_al_2004, Hellinger_al_2011} constitute a less favorable condition for stable small scale magnetic structure as current sheets, as well as planetary magnetosheath where enhanced $T_\perp>T_\|$ anisotropies are often observed for the downstream protons \citep{Samsonov_al_2007}.

The coupling between particle properties and the evolution of small scale magnetic structures, as described in this work, will be more deeply investigated with the help of measurements from the future heliospheric explorations, as Solar Orbiter and Solar Probe Plus, where combined particle and field high resolution observations are planned.

\begin{acknowledgments}
The research leading to these results has received funding from the European Commission's Seventh Framework Programme (FP7/2007-2013) under the grant agreement SHOCK (project number 284515), from the US NSF SHINE program, and from the Science and Technology Facilities Council 
(STFC).
It was also carried out in part at the Jet Propulsion Laboratory, California Institute of Technology, under a contract with the National Aeronautics and Space Administration.
\end{acknowledgments}




\begin{thebibliography}{42}
\expandafter\ifx\csname natexlab\endcsname\relax\def\natexlab#1{#1}\fi

\bibitem[{{Ambrosiano} {et~al.}(1986){Ambrosiano}, {Lee}, \&
  {Fu}}]{Ambrosiano_al_1986}
{Ambrosiano}, J., {Lee}, L.~C., \& {Fu}, Z.~F. 1986, J. Geophys. Res., 91, 113

\bibitem[{{Aunai} {et~al.}(2011){Aunai}, {Belmont}, \& {Smets}}]{Aunai_al_2011}
{Aunai}, N., {Belmont}, G., \& {Smets}, R. 2011, J. Geophys. Res., 116, 9232

\bibitem[{{Bale} {et~al.}(2009){Bale}, {Kasper}, {Howes}, {Quataert}, {Salem},
  \& {Sundkvist}}]{Bale_al_2009}
{Bale}, S.~D., {Kasper}, J.~C., {Howes}, G.~G., {Quataert}, E., {Salem}, C., \&
  {Sundkvist}, D. 2009, Phys. Rev. Lett., 103, 211101

\bibitem[{{Belmont} {et~al.}(2012){Belmont}, {Aunai}, \&
  {Smets}}]{Belmont_al_2012}
{Belmont}, G., {Aunai}, N., \& {Smets}, R. 2012, Physics of Plasmas, 19, 022108

\bibitem[{{Borovsky} \& {Denton}(2011)}]{Borovsky_Denton_2011}
{Borovsky}, J.~E., \& {Denton}, M.~H. 2011, ApJ Lett., 739, L61

\bibitem[{{Chen} \& {Davidson}(1981)}]{Chen_Davidson_1981}
{Chen}, J., \& {Davidson}, R.~C. 1981, Phys. Fluids, 24, 2208

\bibitem[{{Chen} \& {Palmadesso}(1984)}]{Chen_Palmadesso_1984}
{Chen}, J., \& {Palmadesso}, P. 1984, Physics of Fluids, 27, 1198

\bibitem[{{Coppi}(1983)}]{Coppi_1983}
{Coppi}, B. 1983, ApJ Lett., 273, L101

\bibitem[{{Daughton}(1999)}]{Daughton_1999}
{Daughton}, W. 1999, Physics of Plasmas, 6, 1329

\bibitem[{{Davidson} \& {V\"olk}(1968)}]{Davidson_Volk_1968}
{Davidson}, R.~C., \& {V\"olk}, H.~J. 1968, Phys. Fluids, 11, 2259

\bibitem[{{Drake} {et~al.}(2010){Drake}, {Opher}, {Swisdak}, \&
  {Chamoun}}]{Drake_al_2010}
{Drake}, J.~F., {Opher}, M., {Swisdak}, M., \& {Chamoun}, J.~N. 2010, ApJ, 709,
  963

\bibitem[{{Erd{\H o}S} \& {Balogh}(2008)}]{Erdos_Balogh_2008}
{Erd{\H o}S}, G., \& {Balogh}, A. 2008, Adv. Space Res., 41, 287

\bibitem[{{Gary} {et~al.}(2003){Gary}, {Yin}, {Winske}, {Ofman}, {Goldstein},
  \& {Neugebauer}}]{Gary_al_2003}
{Gary}, S.~P., {Yin}, L., {Winske}, D., {Ofman}, L., {Goldstein}, B.~E., \&
  {Neugebauer}, M. 2003, J. Geophys. Res., 108, 1068

\bibitem[{{Gosling}(2007)}]{Gosling_2007}
{Gosling}, J.~T. 2007, ApJ Lett., 671, L73

\bibitem[{{Gosling} {et~al.}(2006){Gosling}, {Eriksson}, {Skoug}, {McComas}, \&
  {Forsyth}}]{Gosling_al_2006}
{Gosling}, J.~T., {Eriksson}, S., {Skoug}, R.~M., {McComas}, D.~J., \&
  {Forsyth}, R.~J. 2006, ApJ, 644, 613

\bibitem[{{Harris}(1962)}]{Harris_1962}
{Harris}, E.~G. 1962, Il Nuovo Cimento, 23

\bibitem[{{Hellinger}(2007)}]{Hellinger_2007}
{Hellinger}, P. 2007, Phys. Plasmas, 14, 2105

\bibitem[{{Hellinger} {et~al.}(2011){Hellinger}, {Matteini}, {{\v
  S}tver{\'a}k}, {Tr{\'a}vn{\'{\i}}{\v c}ek}, \& {Marsch}}]{Hellinger_al_2011}
{Hellinger}, P., {Matteini}, L., {{\v S}tver{\'a}k}, {\v S}.,
  {Tr{\'a}vn{\'{\i}}{\v c}ek}, P.~M., \& {Marsch}, E. 2011, J. Geophys. Res.,
  116, 9105

\bibitem[{{Hellinger} {et~al.}(2006){Hellinger}, {Tr{\'a}vn{\'{\i}}{\v c}ek},
  {Kasper}, \& {Lazarus}}]{Hellinger_al_2006}
{Hellinger}, P., {Tr{\'a}vn{\'{\i}}{\v c}ek}, P., {Kasper}, J.~C., \&
  {Lazarus}, A.~J. 2006, Geophys. Res. Lett., 33, 9101

\bibitem[{{Kunz} {et~al.}(2011){Kunz}, {Schekochihin}, {Cowley}, {Binney}, \&
  {Sanders}}]{Kunz_al_2011}
{Kunz}, M.~W., {Schekochihin}, A.~A., {Cowley}, S.~C., {Binney}, J.~J., \&
  {Sanders}, J.~S. 2011, MNRAS, 410, 2446

\bibitem[{{Landi} {et~al.}(2012){Landi}, {Matteini}, \&
  {Pantellini}}]{Landi_al_2012}
{Landi}, S., {Matteini}, L., \& {Pantellini}, F. 2012, ApJ, 760, 143

\bibitem[{{Marsch} {et~al.}(2004){Marsch}, {Ao}, \& {Tu}}]{Marsch_al_2004}
{Marsch}, E., {Ao}, X.-Z., \& {Tu}, C.-Y. 2004, J. Geophys. Res., 109, 4102

\bibitem[{{Marsch} {et~al.}(1982){Marsch}, {Schwenn}, {Rosenbauer},
  {Muehlhaeuser}, {Pilipp}, \& {Neubauer}}]{Marsch_al_1982a}
{Marsch}, E., {Schwenn}, R., {Rosenbauer}, H., {Muehlhaeuser}, K.-H., {Pilipp},
  W., \& {Neubauer}, F.~M. 1982, J. Geophys. Res., 87, 52

\bibitem[{{Maruca} {et~al.}(2012){Maruca}, {Kasper}, \&
  {Gary}}]{Maruca_al_2012}
{Maruca}, B.~A., {Kasper}, J.~C., \& {Gary}, S.~P. 2012, ApJ, 748, 137

\bibitem[{Matteini {et~al.}(2012)Matteini, Hellinger, Landi, Tr\'avn\'icek, \&
  Velli}]{Matteini_al_2011}
Matteini, L., Hellinger, P., Landi, S., Tr\'avn\'icek, P., \& Velli, M. 2012,
  Space Sci. Rev., 172, 373

\bibitem[{{Matteini} {et~al.}(2007){Matteini}, {Landi}, {Hellinger},
  {Pantellini}, {Maksimovic}, {Velli}, {Goldstein}, \&
  {Marsch}}]{Matteini_al_2007}
{Matteini}, L., {Landi}, S., {Hellinger}, P., {Pantellini}, F., {Maksimovic},
  M., {Velli}, M., {Goldstein}, B.~E., \& {Marsch}, E. 2007, Geophys. Res.
  Lett., 34, 20105

\bibitem[{{Matteini} {et~al.}(2006){Matteini}, {Landi}, {Hellinger}, \&
  {Velli}}]{Matteini_al_2006}
{Matteini}, L., {Landi}, S., {Hellinger}, P., \& {Velli}, M. 2006, J. Geophys.
  Res., 111, 10101

\bibitem[{{Matthews}(1994)}]{Matthews_1994}
{Matthews}, A.~P. 1994, J. Comp. Phys., 112, 102

\bibitem[{{Osman} {et~al.}(2012{\natexlab{a}}){Osman}, {Matthaeus}, {Hnat}, \&
  {Chapman}}]{Osman_al_2012a}
{Osman}, K.~T., {Matthaeus}, W.~H., {Hnat}, B., \& {Chapman}, S.~C.
  2012{\natexlab{a}}, Phys. Rev. Lett., 108, 261103

\bibitem[{{Osman} {et~al.}(2012{\natexlab{b}}){Osman}, {Matthaeus}, {Wan}, \&
  {Rappazzo}}]{Osman_al_2012b}
{Osman}, K.~T., {Matthaeus}, W.~H., {Wan}, M., \& {Rappazzo}, A.~F.
  2012{\natexlab{b}}, Phys. Rev. Lett., 108, 261102

\bibitem[{{Quest} {et~al.}(2010){Quest}, {Karimabadi}, \&
  {Daughton}}]{Quest_al_2010}
{Quest}, K.~B., {Karimabadi}, H., \& {Daughton}, W. 2010, Phys. Plasmas, 17,
  022107

\bibitem[{{Riquelme} {et~al.}(2012){Riquelme}, {Quataert}, {Sharma}, \&
  {Spitkovsky}}]{Riquelme_al_2012}
{Riquelme}, M.~A., {Quataert}, E., {Sharma}, P., \& {Spitkovsky}, A. 2012,
  \apj, 755, 50

\bibitem[{{Samsonov} {et~al.}(2007){Samsonov}, {Alexandrova}, {Lacombe},
  {Maksimovic}, \& {Gary}}]{Samsonov_al_2007}
{Samsonov}, A.~A., {Alexandrova}, O., {Lacombe}, C., {Maksimovic}, M., \&
  {Gary}, S.~P. 2007, Annales Geophysicae, 25, 1157

\bibitem[{{Schekochihin} {et~al.}(2005){Schekochihin}, {Cowley}, {Kulsrud},
  {Hammett}, \& {Sharma}}]{Schekochihin_al_2005}
{Schekochihin}, A.~A., {Cowley}, S.~C., {Kulsrud}, R.~M., {Hammett}, G.~W., \&
  {Sharma}, P. 2005, ApJ, 629, 139

\bibitem[{{Schekochihin} {et~al.}(2010){Schekochihin}, {Cowley}, {Rincon}, \&
  {Rosin}}]{Schekochihin_al_2010}
{Schekochihin}, A.~A., {Cowley}, S.~C., {Rincon}, F., \& {Rosin}, M.~S. 2010,
  MNRAS, 405, 291

\bibitem[{{Schoeffler} {et~al.}(2011){Schoeffler}, {Drake}, \&
  {Swisdak}}]{Schoeffler_al_2011}
{Schoeffler}, K.~M., {Drake}, J.~F., \& {Swisdak}, M. 2011, ApJ, 743, 70

\bibitem[{{Servidio} {et~al.}(2011){Servidio}, {Greco}, {Matthaeus}, {Osman},
  \& {Dmitruk}}]{Servidio_al_2011}
{Servidio}, S., {Greco}, A., {Matthaeus}, W.~H., {Osman}, K.~T., \& {Dmitruk},
  P. 2011, J. Geophys. Res., 116, 9102

\bibitem[{{Servidio} {et~al.}(2012){Servidio}, {Valentini}, {Califano}, \&
  {Veltri}}]{Servidio_al_2012}
{Servidio}, S., {Valentini}, F., {Califano}, F., \& {Veltri}, P. 2012, Physical
  Review Letters, 108, 045001

\bibitem[{{Sharma} {et~al.}(2007){Sharma}, {Quataert}, {Hammett}, \&
  {Stone}}]{Sharma_al_2007}
{Sharma}, P., {Quataert}, E., {Hammett}, G.~W., \& {Stone}, J.~M. 2007, ApJ,
  667, 714

\bibitem[{{Shi} {et~al.}(1987){Shi}, {Lee}, \& {Fu}}]{Shi_al_1987}
{Shi}, Y., {Lee}, L.~C., \& {Fu}, Z.~F. 1987, J. Geophys. Res., 92, 12171

\bibitem[{{{\v S}tver{\'a}k} {et~al.}(2008){{\v S}tver{\'a}k},
  {Tr{\'a}vn{\'{\i}}{\v c}ek}, {Maksimovic}, {Marsch}, {Fazakerley}, \&
  {Scime}}]{Stverak_al_2008}
{{\v S}tver{\'a}k}, {\v S}., {Tr{\'a}vn{\'{\i}}{\v c}ek}, P., {Maksimovic}, M.,
  {Marsch}, E., {Fazakerley}, A.~N., \& {Scime}, E.~E. 2008, J. Geophys. Res.,
  113, 3103

\bibitem[{{Wicks} {et~al.}(2013){Wicks}, {Matteini}, {Horbury}, {Hellinger}, \&
  {Roberts}}]{Wicks_al_2012}
{Wicks}, R.~T., {Matteini}, L., {Horbury}, T.~S., {Hellinger}, P., \&
  {Roberts}, A.~D. 2013, Proc. of the Thirteenth International Solar Wind
  Conference

\end{thebibliography}
\end{document}